"Electrical conductivity characterized at varying strains in spiral cut high-pressure torsion discs"

Evander Ramos[a], Takahiro Masuda[b], Zenji Horita[b], Suveen Mathaudhu[a,c]*
[a]University of California, Riverside, USA
[b]Kyushu University, Fukuoka, Japan
[c]Colorado School of Mines, USA

* Corresponding author: smathaudhu@mines.edu, 1500 Illinois St., Golden, CO 80401, USA

**Abstract**

High-pressure torsion (HPT) imparts inhomogeneous strain to process discs with low strain in the center and higher strain at the outer edge. Microscopy and microhardness indentation have been used to characterize and correlate this inhomogeneity with strain, but similar exploration with other properties has been uncommon. In this work, the electrical conductivity of pure copper discs processed by HPT was characterized with respect to equivalent strain by cutting them into spirals with an incremental, monotonic increase in strain. Electrical conductivity varied with straining in agreement with the literature and expectations based on grain boundary evolution. The spiral conductivity testing method outlined in this work can improve characterization of HPT materials in future studies.

**1. Introduction**

High-pressure torsion (HPT) is a well-documented severe plastic deformation (SPD) technique with the potential to impart large strains, albeit inhomogeneously. This inhomogeneity arises from the torsional straining, which can be described by the equation $\varepsilon = \frac{2\pi r N}{\sqrt{3}\, t}$, where $\varepsilon$=equivalent strain, $r$=radial distance, $N$=number of rotations, and $t$=thickness [1]. Microscopy and microhardness indentation at discrete locations throughout samples have been used to characterize inhomogeneity with processing strain. However, some properties are characterized via tests across large volumes encompassing wide ranges of strains, thereby obfuscating their evolution with processing.

Electrical conductivity in HPT discs has been characterized via four-point probe in tests across cut strips [1–4] or co-linear surface tests [5,6], eddy current techniques [7], and across entire samples [8,9]. Most works report single values for the conductivity of discs, disregarding inhomogeneity across the specific tested regions. Only a few works have explicitly incorporated inhomogeneity in characterizing conductivity across discs [2–5]. However, these tests suffer from inflexibility in probing different strain levels and measurement errors due to small sample sizes. Thus, improved testing methods are necessary for more versatile and accurate strain-based conductivity characterization of HPT samples.

In this work, HPT processed pure copper discs were cut into uniform spiral wires with monotonically increasing equivalent strain to probe conductivity. Measurements between specific strain ranges enabled accurate conductivity characterization across discs. Pure copper was used for its relevance in electrical applications, along with the wealth of SPD literature available for comparison. The results gathered agreed with SPD literature, demonstrating the improved efficiency and precision of this spiral conductivity testing method for characterizing variations with strain in HPT discs.

## 2. Materials and Methods

Discs of high purity copper were processed in a prior report, where microhardness and microstructures at varying strains were also characterized [10]. In this work, an electrical discharge machine was used to cut these discs into an Archimedean spiral shape with a uniform 0.6 mm square cross section. This shape facilitated electrical conductivity characterization using a four-point probe technique with a Keithley 2636B SourceMeter. As illustrated in Figure 1a, two DC source leads were placed at the center and outer section of the spiral and two voltage sensing leads were placed at varying points along the spiral to measure the response at different sections (colored red, blue, and green) as well as for the entire length. Tests used a current of 1 Amp at room temperature. The current density applied during these tests was four orders of magnitude below the threshold for electromigration, so microstructural changes during testing was negligible [11]. Current reversal and line cycle integration were used to cancel out thermoelectric electromotive force and power-line noise, respectively, as recommended for low-voltage measurements [12]. Resistivity was determined from the uniform cross-sectional area and the length between probes. Error propagation was calculated as the root sum squared combination of the measurement uncertainty of the wire thickness, width, and length, as well as the uncertainty between five measurements for each section. Due to physical constraints necessary to cut the spirals, the centermost and outermost regions of the discs were excluded from characterization.

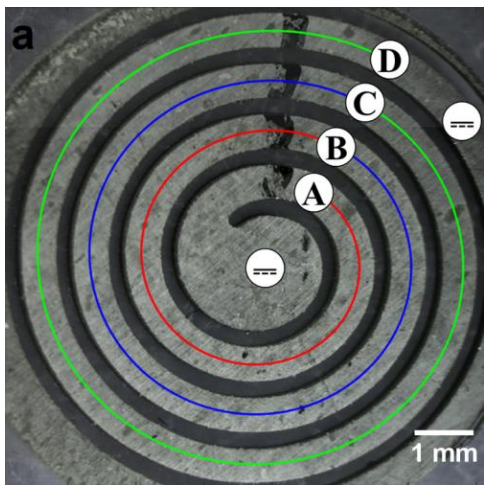
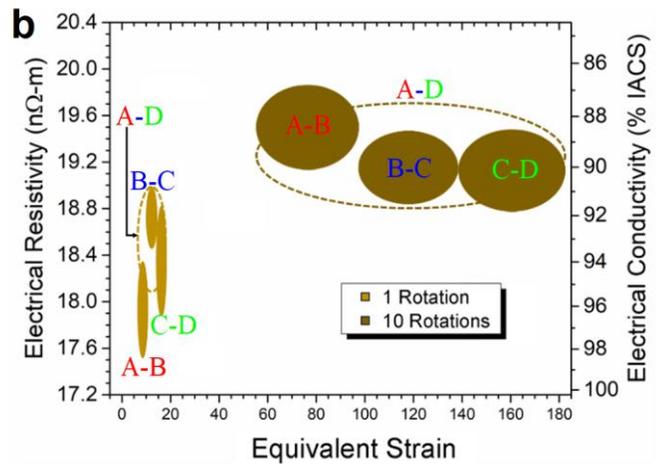

Figure 1: (a) Spiral cut HPT disc overlaid with the locations of the current source probes (=) and voltage sense leads (A-D). (b) Electrical resistivity and conductivity variation with equivalent strain for the three spiral sections (solid filled) and the entire spiral (dashed) for 1 and 10-rotation discs. The y-spread is the standard error of the measurements, while the x-spread is the range of equivalent strains covered by the tested section.

**3. Results and discussion**

3.1. Spiral conductivity testing method results

The results of the conductivity tests are shown in Figure 1b, with the conductivity of pure copper, nominally 100% IACS (International Annealed Copper Standard), near the x-axis. The conductivities for the entire spirals (A-D) of the 1 and 10-rotation discs were 93.2±2.1% and 89.5±2.1% IACS, respectively. Combining the individual measurements from the three spiral sections gave 93.8% and 89.7% IACS for the 1 and 10-rotation discs, respectively, corresponding well with the entire spirals. Generally, the entire spiral covers a wide range of equivalent strains, much like in prior reports testing conductivity across cut HPT strips or entire discs. The high conductivity measured for the innermost tested 1-rotation region (A-B) lies outside the error bars for the entire spiral, so essentially, had traditional conductivity testing been used, this behavior would not have been observed. The 1-rotation sample had an average grain size of 700±200 nm, while the 10-rotation sample had an average of 290±150 nm, as shown previously [10] and in agreement with published literature [1]. Using an approximation of grain boundary contribution to electrical resistance in copper at room temperature [13], resistance should increase by about 1.0 nΩ-m and 2.5 nΩ-m for the 1 and 10-rotation discs, respectively, in line with the values presented in Figure 1b.

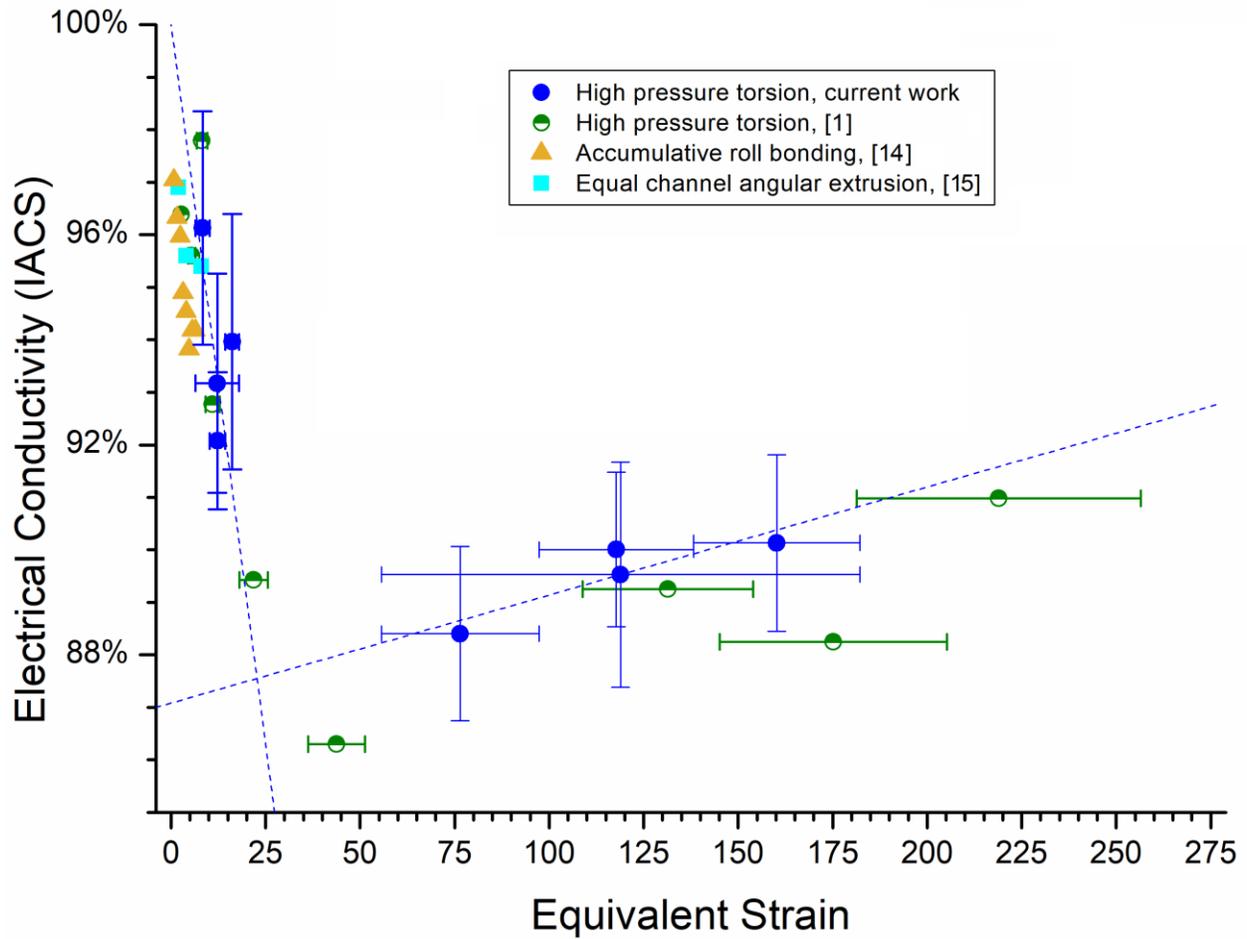

Figure 2. Electrical conductivity variation with equivalent strain in the present work and other SPD studies. All works [1,14,15] used 99.99% Cu. Trendlines for the two discs in the current work are also included.

3.2. Comparison to SPD literature

The present results were compared across the SPD copper literature in Figure 2. Prior works measuring conductivity after cycles of accumulative roll bonding [14] and equal channel angular extrusion [15] agreed with the measurements from the 1-rotation disc. Another HPT work conducted four-point probe testing on strips cut from many discs, which agreed with the values measured presently from just two discs. Linear extrapolation of the results from these two discs gave an inflection point for conductivity at an equivalent strain around 25, with a minimum value of 87.5% IACS. This agrees with the literature, where a strain of around 20 has been cited as the onset of saturation in hardness and microstructural evolution in copper [1,10]. Although the 10-rotation disc error bars overlap, they show a slight increase in conductivity with equivalent strain, which was also indicated in the prior work [1], so the limits of this conductivity evolution at high strains may require more exploration in the future.

3.3. Comparison to other strain-based testing methods

Only a few works have characterized conductivity with strain throughout individual HPT samples. Another Cu work [3] used a four-point probe method on small sections cut from disc diameters, but they also measured grain boundary length at varying strains and saw no correlation with conductivity. One explanation for this could be the measurement procedure itself. With characterization conducted on 1 mm strips, inaccuracy in measuring such a short length would potentially contribute a large percentage of error for resistivity calculations. In the present work, the shortest spiral section was much longer, and exhaustive efforts were implemented to mitigate and calculate the propagation of error. Another prior work used colinear four-point probe testing to map conductivity throughout titanium and zirconium discs [5]. Unfortunately, variation in geometric correction factors with probe placement adds variability to such mapping, especially near the sample edge. In another work, ring samples processed by HPT were cut concentrically like a bullseye to characterize conductivity at three short ranges of equivalent strain [4]. Another recent work explored thermoelectric properties throughout large HPT samples using a few small cut strips [2]. The testing methods used in the present work allow for greater flexibility to probe across broader strain ranges than in these two works. Thus, compared to prior literature characterizing conductivity in different regions of HPT samples, the spiral conductivity testing method used in the present work provides an accurate and efficient framework for exploring electrical behavior with strain.

## 4. Conclusion

Electrical conductivity in copper discs processed by HPT was probed with respect to equivalent strain through spiral cutting and four-point probe testing. Measurements gathered from 1 and 10-rotation discs agree with values from the literature for copper after various SPD processes and conditions. The spiral conductivity testing method showcased in this work can improve assessment of processing-structure-property relationships in other materials, such as alloys, nanocomposites, and thermoelectrics.

**Acknowledgements**

HPT was carried out in the International Research Center on Giant Straining for Advanced Materials (IRC-GSAM) at Kyushu University, Japan. EHR was supported by the GAANN Fellowship from the Department of Education through the Mechanical Engineering department at the University of California, Riverside. SNM was supported via NSF CMMI Grant #1663522.